\documentclass[a4paper]{spie}  

\usepackage{amsmath,amsfonts,amssymb}
\usepackage{graphicx}
\usepackage[colorlinks=true, allcolors=blue]{hyperref}

\usepackage{textcomp} 
\usepackage{xcolor}    

\title{A telescope control and scheduling system for the
Gravitational-wave Optical Transient Observer (GOTO)}

\author[a]{Martin J. Dyer}
\author[a,b]{Vik S. Dhillon}
\author[a]{Stuart Littlefair}
\author[c]{Danny Steeghs}
\author[c]{Krzysztof Ulaczyk}
\author[c]{Paul Chote}
\author[d]{Duncan Galloway}
\author[d]{Evert Rol}
\author[ ]{the GOTO Collaboration}

\affil[a]{Department of Physics and Astronomy, University of Sheffield, Sheffield S3 7RH, UK}
\affil[b]{Instituto de Astrof\'{i}sica de Canarias, E-38205 La Laguna, Tenerife, Spain}
\affil[c]{Department of Physics, University of Warwick, Coventry CV4 7AL, UK}
\affil[d]{Monash Centre for Astrophysics, School of Physics \& Astronomy, Monash University, Clayton VIC 3800, Australia}

\authorinfo{Send correspondence to MJD - email: martin.dyer@sheffield.ac.uk}

\begin{document} 
\maketitle

\begin{abstract}
The Gravitational-wave Optical Transient Observer (GOTO) is a wide-field telescope project aimed at detecting optical counterparts to gravitational wave sources. The prototype instrument was inaugurated in July 2017 on La Palma in the Canary Islands. We describe the GOTO Telescope Control System (G-TeCS), a custom robotic control system written in Python which autonomously manages the telescope hardware and nightly operations. The system comprises of multiple independent control daemons, which are supervised by a master control program known as the ``pilot''. Observations are decided by a ``just-in-time'' scheduler, which instructs the pilot what to observe in real time and provides quick follow-up of transient events.
\end{abstract}


\section{The GOTO Telescope}
\label{sec:goto}

In 2016 the first direct detection of gravitational waves was announced by the LIGO collaboration \cite{PhysRevLett.116.061102}, and despite a global follow-up search no counterpart electromagnetic transient was detected \cite{2016ApJ...826L..13A}. This was not unexpected due to the signal's predicted origin as a binary black hole system. It was not until nearly two years and four more confirmed detections later that the first gravitational waves were detected from a merging binary neutron star \cite{PhysRevLett.119.161101}. This GW170817 LIGO/Virgo detection marked a milestone in the era of multi-messenger astronomy, as it was also seen in gamma-rays as GRB 170817A by the \textit{Fermi} satellite and 11 hours later as optical transient AT 2017gfo \cite{2017ApJ...848L..12A}. GW170817 was localised to a 90\% confidence interval covering 31 square degrees \cite{2017ApJ...848L..12A}, lower than the typical expected areas of hundreds of square degrees \cite{0004-637X-795-2-105}, which made prompt discovery of the associated transient possible for telescopes with a small field of view. Future detections may not be as well localised and therefore in order to ensure future counterpart observations there will be a need for wide-field searches.

\begin{figure} [htb]
\begin{center}
\begin{tabular}{c}
\includegraphics[width=11cm]{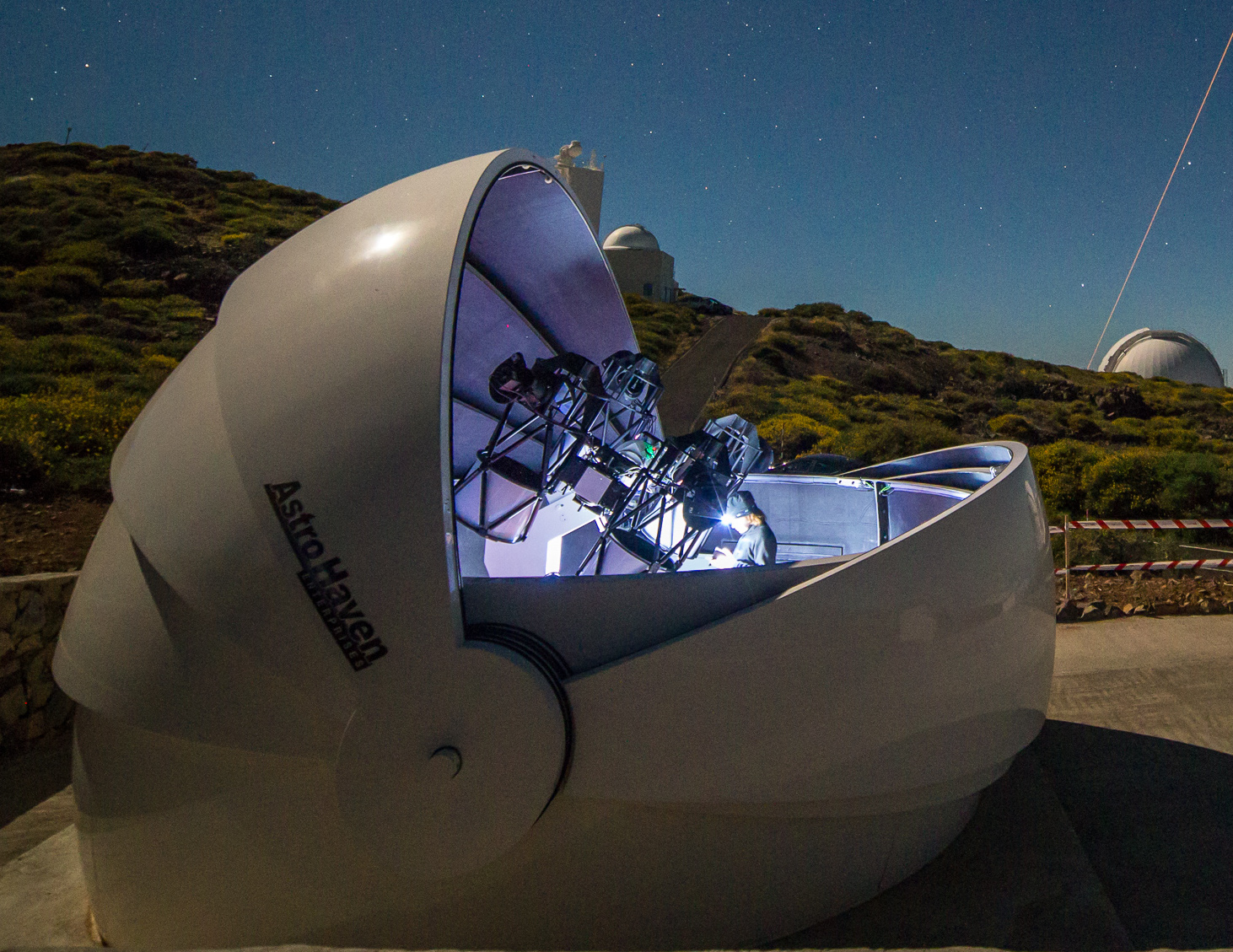}
\end{tabular}
\end{center}
\caption[example] 
{ \label{fig:goto_photo} 
The GOTO prototype instrument on La Palma, with four of the eventual eight unit telescopes.}
\end{figure}

The Gravitational-wave Optical Transient Observer (GOTO) is a project dedicated to detecting future optical counterparts of gravitational wave sources by employing a wide-field approach to be able to localize such counterparts early\cite{GOTO-website}. The first prototype instrument was inaugurated at the Roque de los Muchachos Observatory on La Palma, Canary Islands in July 2017 and is shown in Fig.~\ref{fig:goto_photo}. GOTO uses arrays of 40~cm unit telescopes (UTs) on a single fast-slewing mount. This modular approach allows the project to scale to large fields of view in a cost-effective manner. A full instrument will have 8 of these UTs per mount, giving an overall field of view of 40 square degrees with a pixel scale of 1.2 arcsec/pix and a limiting magnitude of $\sim$20 in each two minute exposure. Each UT has a set of wide white and coloured filters to assist source characterisation.  An additional instrument with 8 more UTs on a second mount is planned to be co-located in a second dome on La Palma, which will double the instantaneous field of view to 80 square degrees, allow the sky to be surveyed at a higher cadence and give more options for transient follow-up. A southern node is also planned for Australia.

\begin{figure} [htb]
\begin{center}
\begin{tabular}{c}
\includegraphics[width=11cm]{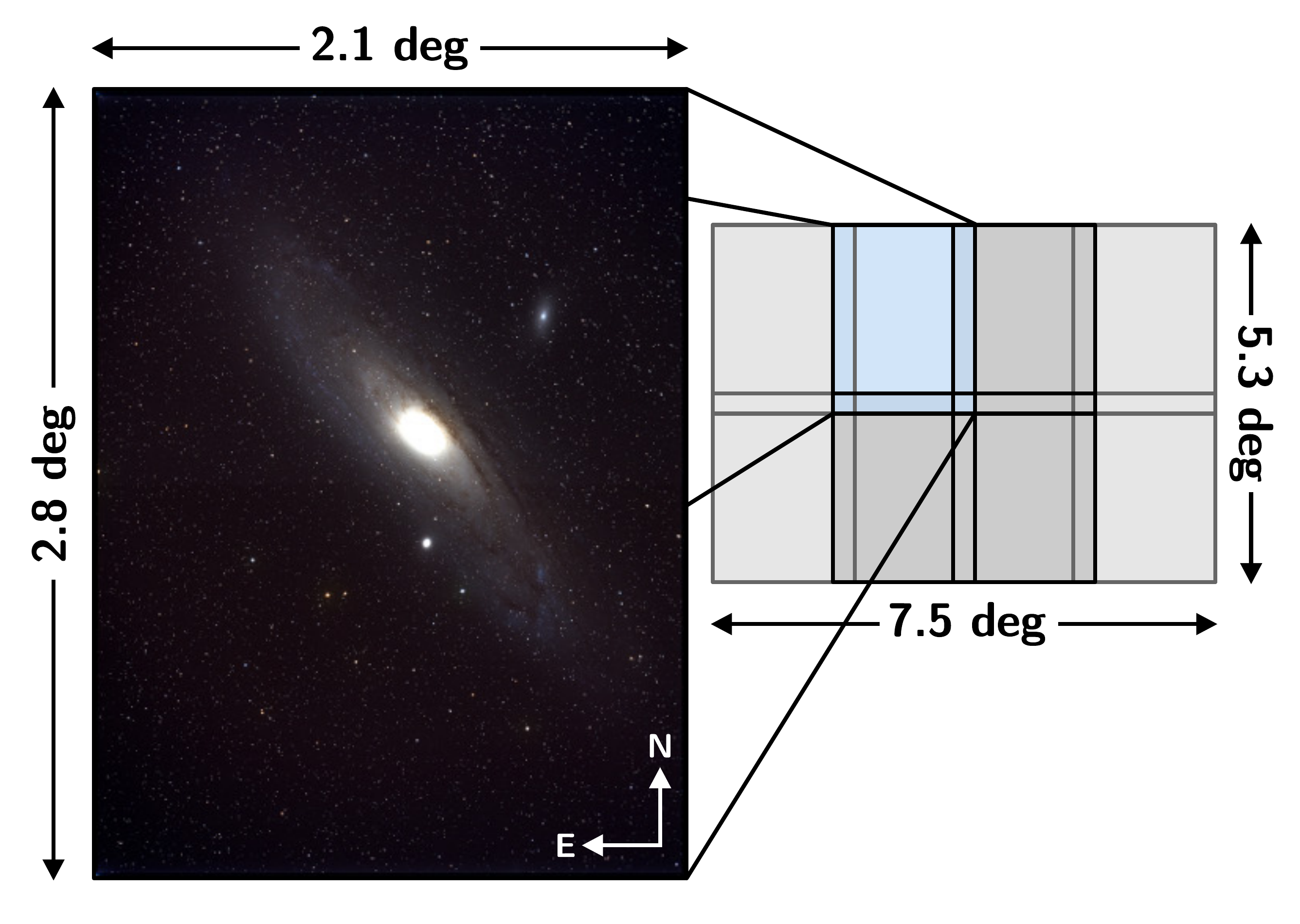}
\end{tabular}
\end{center}
\caption[example] 
{ \label{fig:tiles} 
A single commissioning image of M31, showing the wide field of view of each unit telescope. The full 8 unit telescope array will cover an area of approximately 40 square degrees, which forms a single survey tile.}
\end{figure}

GOTO is designed as a robotic, wide-field transient detection telescope. Under normal circumstances the telescope will carry out an all-sky survey, based on a fixed grid of tiles. Each tile corresponds to the 40 square degree field of view of the 8 UTs observing neighbouring patches of sky, as shown in Fig.~\ref{fig:tiles}. The all-sky survey will aim to obtain a high-cadence at an appropriate depth, mapping the entire visible sky several times a week. The high cadence will ensure that there are very recent reference images available to allow detection of objects using difference imaging.


\begin{figure} [ht]
\begin{center}
\begin{tabular}{c}
\includegraphics[width=13cm]{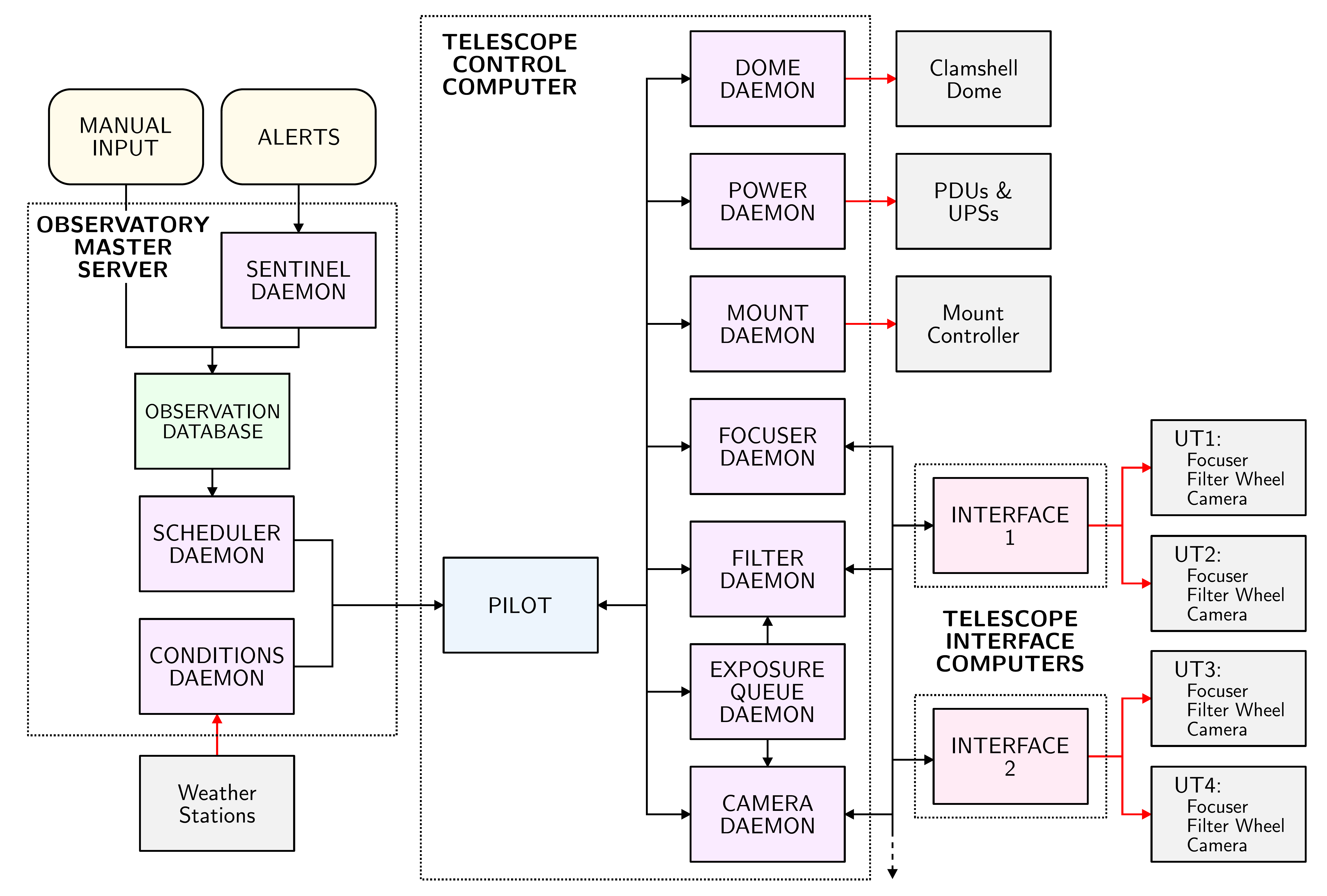}
\end{tabular}
\end{center}
\caption[example] 
{ \label{fig:flow} 
\textsf{G-TeCS} system architecture. The observation database as well as the sentinel, scheduler and conditions daemons shown to the left run on an observatory-wide master server, while the pilot and hardware daemons are located on the telescope control computer within the dome. Control for the unit telescope hardware (focuser, filter wheel and camera) is sent via an interface daemon for each pair of UTs, running on computers located on the mount. Only the system for the prototype instrument (one mount with four unit telescopes) is shown, see Fig.~\ref{fig:flow2} for the proposed full system.}
\end{figure}

\section{Telescope Control}
\label{sec:control}

In order to meet the requirements for controlling GOTO, it was decided to implement a custom telescope control system. The starting point was the control system already developed for \textit{pt5m}, a 0.5m robotic telescope built and operated by Sheffield and Durham Universities and also located on La Palma \cite{2015MNRAS.454.4316H}. The pt5m control system is written in Python and was built around multiple independent hardware daemon programs with a pilot program to control them. This formed the basis for the \textit{GOTO Telescope Control System} (\textsf{G-TeCS}), which built upon the pt5m system to adapt and upgrade it specifically for use on the GOTO project. The core \textsf{G-TeCS} system architecture is shown in Fig.~\ref{fig:flow} and is described in the following sections.

\begin{figure} [ht]
\begin{center}
\begin{tabular}{c}
\includegraphics[width=17cm]{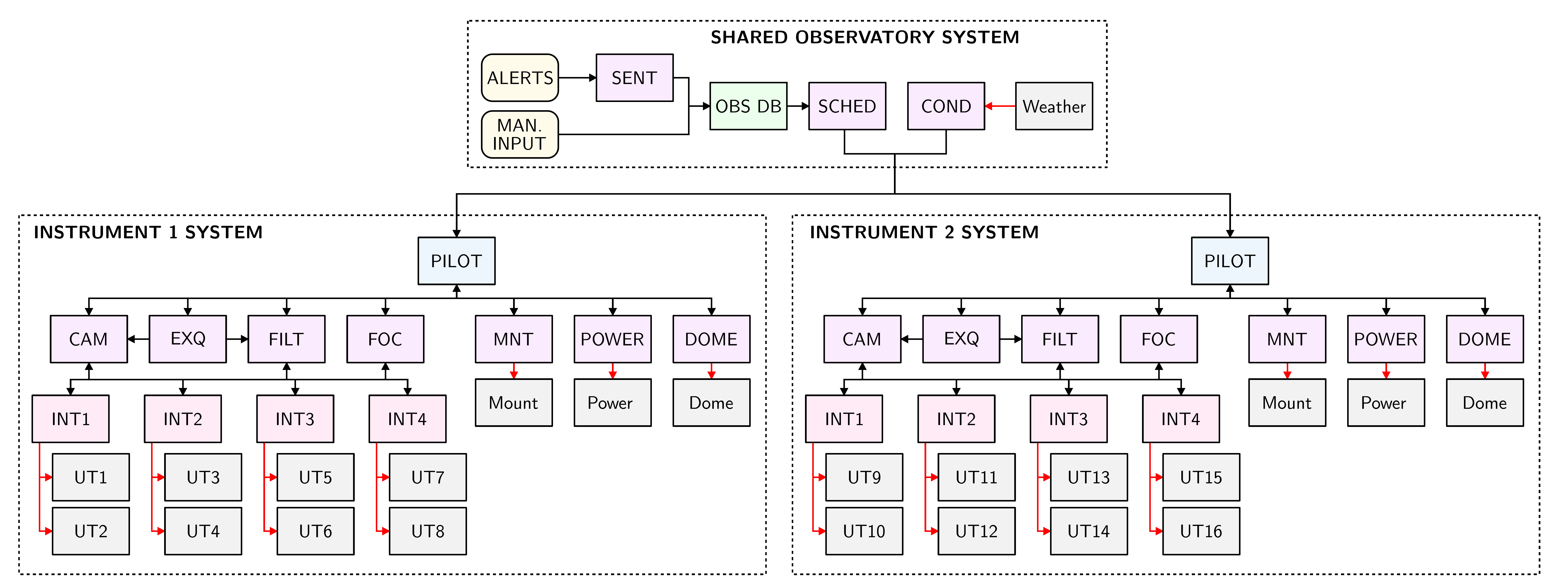}
\end{tabular}
\end{center}
\caption[example] 
{ \label{fig:flow2} 
An expanded version of Fig.~\ref{fig:flow}, showing the proposed full system architecture for two GOTO mounts each with eight unit telescopes. Each telescope is envisioned to have an independent pilot while sharing common scheduling and conditions monitoring systems running on the central server.}
\end{figure}

\renewcommand{\thefootnote}{\alph{footnote}}

\subsection{Hardware Control}
\label{sec:hardware}

The core elements of the control system are the hardware daemons. A \textit{daemon} is a type of computer program that runs as a background process, continually cycling and awaiting any input from the user. This makes it ideal for hardware control as each daemon can constantly monitor its associated hardware and then execute commands when received via a command line script. Each category of hardware has a dedicated control daemon, meaning that while the dome daemon only controls the single dome the camera daemon centralises commands to all cameras on the one mount (8 in the full system).

Each daemon is built around a Python class which contains hardware control functions and a main loop. The loop is set running in its own thread, and when a control function is called it issues the relevant command to the hardware. The daemons are created using the Python Remote Objects (\textsf{Pyro}) module\cite{Pyro-website}. Each daemon is a \textsf{Pyro} server, any client script can then access its functions across the network using the associated server ID. For ease of use each daemon has an associated control script, which allows commands run by a user in a terminal to be executed by the daemon. In robotic mode these commands will be issued by the pilot (see Sec~\ref{sec:pilot}). Commands follow a simple format, e.g. \texttt{foc~home} homes all connected focusers, \texttt{dome~close} closes the dome, \texttt{cam~image~2~60} will take a 60 second science exposure on camera 2 only.

There are seven primary hardware control daemons, as shown in Fig.~\ref{fig:flow}. A brief description of each is given in the following sections.

\subsubsection{Dome}
\label{sec:dome}
The dome daemon (command name \texttt{dome}) sends commands to the Astrohaven clamshell dome\footnote{Astro Haven Enterprises, CA, USA; (\url{www.astrohaven.com})} through a serial connection. It also independently monitors the output of the conditions daemon, meaning it will close the dome if any of the conditions flags are set to bad (see Sec.~\ref{sec:conditions}). As a safety feature it will also refuse to open if a \texttt{dome open} command is issued while external conditions are bad, although this can be overridden when in manual mode (see Sec.~\ref{sec:pilot}). The dome daemon will also automatically turn on the dome dehumidifier if the internal humidity gets too high, and will turn it off when it reaches a lower level or if the dome is opened.

The GOTO dome also has two additional Arduino boards fitted inside. The first is connected to additional sensors attached to the dome shutters and hatch door to act as backups to the in-built open/close sensors, as well as a siren which the dome daemon will sound whenever the dome is about to move. The second Arduino acts as an emergency backup system in case the dome daemon or control computer is disabled. The dome daemon emits a heartbeat signal to this board every second, and if this signal is not received after ten seconds the Arduino will take over the serial connection and issue commands to close the dome.

\subsubsection{Power}
\label{sec:power}
The power daemon (\texttt{power}) provides a unified interface for the various networked power units within the GOTO dome. This includes two uninterruptible power supplies (UPSs), one for the mount and electronics rack and a second for the dome, the power distribution units (PDUs) in the rack which power the individual computers and units, and the DC switch boards on the mount boom arm which power the interface computers and camera hardware (see Sec.~\ref{sec:focfiltcam}). Each power outlet has a unique name assigned and can be powered on or off through the power daemon, this gives the ability to remotely power cycle each system on the network in case of any problems. Similar outlets can be grouped together even if they are on different units, so the command \texttt{power on cams} will power on all of the CCD cameras. 

\newpage

\subsubsection{Mount}
\label{sec:mnt}
The mount daemon (\texttt{mnt}, so as not to interfere with the Linux \texttt{mount} command) sends commands to the GOTO mount. The mount hardware is connected to a SiTech servo controller, and the system uses the SiTech control software\footnote{Sidereal Technology, OR, USA; (\url{www.siderealtechnology.com})}. This is run as \textsf{SiTechEXE} on a Windows computer attached to the mount. Initially commands could only be sent through the \textsf{ASCOM} system running on that computer, which required an interface daemon to convert between \textsf{ASCOM} and Python commands accessible using \textsf{Pyro}. However an update to \textsf{SiTechEXE} added the ability to send TCP commands, meaning there is no need for an interface and now the mount daemon can communicate with it directly. Using the SiTech software provides several advantages, in particular an inbuilt pointing model system using \textsf{PointXP}.

\subsubsection{Focusers / Filter wheels / Cameras}
\label{sec:focfiltcam}
The GOTO detectors use off-the-shelf hardware from FLI\footnote{Fingerlake Instruments, NY, USA; (\url{www.flicamera.com})}: each GOTO unit telescope uses a MicroLine ML50100 camera with a KAF-50100 CCD detector, an Atlas focuser and a CFW9-5 filter wheel with a set of Baader LRGBC filters\footnote{Baader Planetarium, Germany; (\url{www.baader-planetarium.com})}. A GOTO mount will hold up to eight unit telescopes each equipped with a camera, focuser and filter wheel. Each piece of hardware needs to be connected by USB, but running 24 USB cables down the mount was deemed impractical. Therefore for each pair of UTs these cables are connected to a small PC - an Intel NUC in a sealed fanless enclosure - attached to the boom arm. Interface scripts are run on these, as shown in Fig.~\ref{fig:flow}. Using the FLI SDK a Cython wrapper called \textsf{fli-api} was written to allow easy control through Python, and using \textsf{Pyro} these functions are published to the network to be accessed by the control daemons on the main control computer. This allows all the pieces of the hardware to be controlled individually.

The camera (\texttt{cam}), focuser (\texttt{foc}) and filter (\texttt{filt}) daemons each issue commands to every one of their connected pieces of hardware. Commands can be sent independently or to all: \texttt{cam~image~2~60} will take a 60 second exposure on camera 2 only while not including the specifier (\texttt{cam~image~60}) would execute it on all cameras simultaneously. Multiple selections can be made using a simple comma-separated syntax, such as \texttt{cam~image~1,2,4~60}.

There is also a fourth related daemon called the exposure queue daemon (\texttt{exq}), which coordinates taking frames in sequence and setting filters before the exposures start. The command \texttt{exq~image~60~L} would add a single exposure in the L filter to the queue while \texttt{exq~multiimage~3~60~L} would add three. The command \texttt{exq~image~60~R,G,B} would add three images, each to be taken in the R, G and B filters, and the exposure queue daemon would issue commands to the filter daemon to move the filter wheel between exposure commands to the camera daemon.

Images are written to FITS files by the camera daemon and are archived by date. Each camera output is saved as a separate file (e.g. \texttt{r00033\_UT2.fits} is the image for run 33 from camera 2). These images are then copied via a dedicated fibre link from La Palma to Warwick University where the photometry pipeline is run.

\renewcommand{\thefootnote}{\fnsymbol{footnote}}

\subsection{Autonomous Observing}
\label{sec:auto}

The hardware control systems described in Sec.~\ref{sec:hardware} provide the basic methods to control the telescope. A human observer could run through a series of simple commands to open the dome, slew the mount, take exposures and repeat for the rest of the night.

GOTO however is designed as a robotic installation, and therefore requires another level of software to take the place as the source of these commands. In \textsf{G-TeCS} this role is filled by a master control program called the pilot. There are also several support daemons that perform roles allowing the system to operate robotically: the conditions daemon which monitors weather and other system conditions, the sentinel daemon which listens for alerts and enters new targets into the observation database and the scheduler daemon which reads that database and calculates what to observe. Each of these systems are described in the following sections.

\newpage

\subsubsection{Pilot}
\label{sec:pilot}

The pilot is a Python script that is run once each night: starting automatically in the late afternoon, running through to the morning and exiting once the system has shut down. It issues commands to the hardware in exactly the same way as a human would, executing the terminal commands which call the control scripts and send the commands to the daemon. The pilot is written as an asynchronous program using the standard Python \textsf{asyncio} module. The program runs with a single-threaded event loop with multiple running tasks as coroutines. Each task contains a loop which executes particular commands and then pauses using the \texttt{await} command to allow other routines to be run. There are two types of routines: the night marshal which runs through tasks as the night progresses and the check routines which monitor the rest of the system. 

Most of the coroutines are designed as monitoring systems to regularly check different parts of the system, which fits well into the asynchronous model. The pilot has two modes of operation: normally it is in robotic mode meaning it is in complete control, but it can be switched into manual mode by a human operator which will suspend the check routines. These check routines are as follows:

\begin{itemize}

\item \texttt{check\_hardware} monitors the hardware daemons, checking every 60 seconds that they are all reporting their expected statuses. If any abnormal status is detected then the pilot is paused (see below) and a series of pre-set recovery commands are executed in turn. When in recovery mode it will check the daemon statuses more frequently. If the commands work and the status returns to normal the pilot is resumed, but if the commands are exhausted without the problem being fixed then an alert is issued (see Sec.~\ref{sec:conditions}) that the system requires human intervention. 

\item \texttt{check\_dome} is a backup to the primary hardware check routine. \texttt{check\_hardware} does monitor the dome along with the other hardware daemons, but \texttt{check\_dome} provides a simple, dedicated backup to ensure the dome is told to close when needed and to raise the alarm if it does not.

\item \texttt{check\_flags} is a routine that monitors the system flags, most notably those created by the conditions monitor (see Sec.~\ref{sec:conditions}). If any of the conditions flags are bad then the dome daemon will close the dome immediately (see Sec.~\ref{sec:dome}) without needing to wait for the pilot. When this happens the \texttt{check\_flags} routine will check that the dome is closed and then pause the pilot and ensure it is not resumed until the flag is cleared. When it is the pilot will reopen the dome and allow normal operations to be restored. The \texttt{check\_flags} routine also monitors the pilot mode flag and will pause the pilot if it is set to manual mode, which suspends the other check routines and any observation scripts launched by the night marshal (see below).

\item \texttt{check\_scheduler} is a routine that queries the scheduler daemon every 10 seconds to find the best job to observe. Each pointing has an ID assigned to its entry in the observation database, as described in Sec.~\ref{sec:obsdb}. It the pilot is currently observing the scheduler will either return the database ID of the current pointing, in which case the pilot will continue with the current job, or a new ID which will lead to the pilot interrupting the current job and moving to observe the new one. The details of how the scheduler decides which target to observe are given in Sec.~\ref{sec:scheduler}.

\end{itemize}

The main pilot nightly operations are carried out by the \texttt{night\_marshal}. This is a separate coroutine which runs through a list of set tasks as the night progresses, based on the altitude of the Sun. Each key task is contained in a separate script, called an observation script, which contains the commands to send to the hardware daemons. These scripts are self-contained processes which mean they can also be called independently, for example if the pilot is not running and a manual observer wants to run the autofocus routine they can use the command \texttt{obs\_script~autoFocus}.  In the order they are performed during the night, the night marshal tasks are:

\begin{enumerate}

\item STARTUP, run immediately when the pilot starts. The night marshal executes the \texttt{startup} script which powers on the camera hardware, unparks the mount, homes the filter wheels and cools the CCDs down to their operating temperature of -20\textdegree C.

\item DARKS, run after the system is started up. This executes the \texttt{takeBiasesAndDarks} script to take a set number of bias and dark frames at the start of the night, usually nine of each. 

\item OPEN, run once the Sun reaches 0{\textdegree} altitude. It simply executes the \texttt{dome~open} command. If the pilot is paused due to bad weather or a hardware fault then the night marshal will wait and not open until the weather improves or fault is fixed. If it is never resolved then the night marshal will remain here until the end of the night and the shutdown timer runs out (see below).

\item FLATS, run once the dome is open and the Sun reaches -1{\textdegree}. This executes the \texttt{takeFlats} script, which moves the telescope into the stored flat position. This is defined as pointing at an altitude of 75{\textdegree} and an azimuth 180{\textdegree} from the sun, using the strategy for wide-field telescopes suggested by the Antarctic Survey Telescope \cite{2014SPIE.9149E..2HW}. The script then takes flat fields in each filter, stepping in position between exposure and automatically increasing the exposure time as the sky darkens.

\item FOCUS, run once the Sun reaches -11{\textdegree}. This executes the \texttt{autoFocus} script, which finds the best focus position by stepping the focusers, measuring the average stellar half-flux diameter (HFD) and finding the minimum based on the known constant ``V-curve'' relationship between focuser position and HFD for each individual unit telescope\cite{weber2001fast}. If the autofocus routine fails for any reason the previous nights' focus positions are restored.

\item OBS, begun once the Sun reaches -15{\textdegree} and continuing for the majority of the night until the Sun reaches -15{\textdegree} again in the morning. When a database ID is received from the scheduler via the \texttt{check\_schedule} routine the \texttt{observe} script is executed. The script queries the observation database (see Sec.~\ref{sec:obsdb}) to get the coordinates and exposure settings for that pointing and then sends the commands to the mount and exposure queue daemons. Once a job is finished, either through completing all its exposures or being interrupted, the entry in the database status entry is updated and the routine awaits the next job from the scheduler.

\item FLATS is repeated once the Sun reaches -10{\textdegree} in the morning, using the same script but this time increasing the exposure times as the sky brightens.

\end{enumerate}

Once the night marshal has completed all of its tasks it exits. The pilot has an internal shutdown timer which set at the beginning of the night to when the rising Sun will reach 0{\textdegree} altitude again. When this time is reached the pilot then runs the \texttt{shutdown} script, which powers off the cameras, parks the mount and ensures the dome is closed. This is on a separate timer to the rest of the night marshal tasks, meaning if something goes wrong during the night the system will always shut down in the morning. It is also possible for the system to emergency shut down during the night, which will occur if a critical error occurs that can not be solved without human intervention (see the definition of a critical conditions flag in Sec.~\ref{sec:conditions}). This ends the pilot for the night, and it falls on whoever checks the alert to restart it if they decide the cause of the alert is resolved.

\subsubsection{Conditions}
\label{sec:conditions}

The conditions daemon is a support daemon that runs on the central server. It takes in readings from the three local weather stations next to the GOTO dome on La Palma, as well as other sources such as internal sensors, every 10 seconds. The daemon processes these inputs into a series of output flags, which have a value of 0 (Good), 1 (Bad) or 2 (ERROR). If any of the flags are marked as not Good then the overall conditions are bad: the dome will close (Sec.~\ref{sec:dome}) and the pilot will pause (Sec.~\ref{sec:pilot}). The conditions daemon is run centrally on the main server because it deals with site-wide values, so when the second instrument is built it is envisioned that they will both share the same conditions daemon (as shown in Fig.~\ref{fig:flow2}).

Each conditions flag has a limit below or above which the flag will turn from good to bad. For categories with multiple sources (for example each of the three weather stations gives an independent external temperature reading) then the limit will be applied to each and if any are found to be bad then the flag is set. Each category also has two parameters, the bad delay and the good delay. These are the time the conditions daemon waits between an input going bad/good and setting the flag accordingly, which has the effect of smoothing out any sudden spikes in a value and ensures the dome will not be opening and closing too often. Certain flags are also marked as critical flags, for conditions that will not be solved by the system pausing and will need human intervention to fix. Alerts are sent out in the form of automated messages to the system Slack channel, with further email and text messages being considered before the system moves to fully unsupervised operations. Alerts are only sent out when critical flags are set, which is why some flags have a normal and a more serious critical version.

An explanation of all the conditions flags is given below. Note the exact values of limits and delay times are often subject to adjustment as commissioning occurs in different weather conditions and seasons.

\begin{itemize}
\item \texttt{rain}: A very simple flag, it is set to bad immediately if any of the weather stations report rain, and will only be cleared after 10 minutes of no more rain reported. In practice rain usually coincides with high humidity, meaning the \texttt{rain} and \texttt{humidity} flags often overlap.

\item \texttt{humidity}: The critical humidity limit is 75\%, with a bad delay of two minutes and a good delay of five minutes.

\item \texttt{temperature} and \texttt{ice}: The \texttt{temperature} flag is set if the temperature drops below 1\textdegree C for two minutes, and has a good delay of five minutes. The \texttt{ice} flag uses the same input but is set if the temperature is below 0\textdegree C for one hour and will only clear if the temperature is above freezing for 24 hours. The \texttt{ice} flag is a critical flag, meaning it will prevent the system from opening until manually cleared (in this case the dome has been checked for any ice formation). 

\item \texttt{windspeed}: Another simple flag that gets set if the windspeed is above 40 km/h, with the same two minute bad delay and five minutes good delay as the other external flags.

\item \texttt{internal}: A combination flag for the two internal temperature and humidity sensors within the dome. These have very extreme limits, a humidity of above 90\% or a temperature of below -3\textdegree C, which should never be reached under normal circumstances due to the internal dehumidifier. This flag therefore is a backup for an emergency case, when either the dehumidifier is not working or the dome has somehow opened in bad conditions. This is a critical flag, in order to alert that something is wrong inside the dome. 

\item \texttt{link}: The conditions daemon also monitors the external internet link to the site, by pinging the Warwick server and other public internet sites. If these pings are unsuccessful after a minute then the \texttt{link} flag gets set to bad. It is technically possible for the system to observe without any internet link, but it is an unnecessary risk as if anything went wrong alerts could not be sent out and external users would not be able to log in.

\item \texttt{diskspace}: The amount of free disk space on the image data drive is also monitored, with the system shutting down and sending out an alert if there is less than 5\% space available. This is a critical flag, as currently space needs to be cleared manually.

\item \texttt{ups} and \texttt{low\_battery}: The conditions daemon will set the \texttt{ups} flag if the observatory has lost power and the system UPSs are discharging. Brief power cuts do occur on La Palma, but rarely for more than a few minutes. The \texttt{low\_battery} flag is a critical flag that gets triggered if the UPSs ever reach less than 75\% battery remaining, which would be a sign of a more serious problem that is worth alerting and forcing a shutdown until it is cleared. 

\item \texttt{hatch}: A critical flag to detect if the access hatch into the dome has been left open. This flag is unique in that is is only valid in robotic mode; when in manual mode it is assumed that the hatch being opened is a result of someone operating the telescope. However once the system is observing robotically the hatch being open is a critical problem, as there is no way to close it remotely and in the case of bad weather damage could be caused to the telescope.

\end{itemize}

\subsubsection{Observation database}
\label{sec:obsdb}

\begin{figure} [ht]
\begin{center}
\begin{tabular}{c}
\includegraphics[width=9cm]{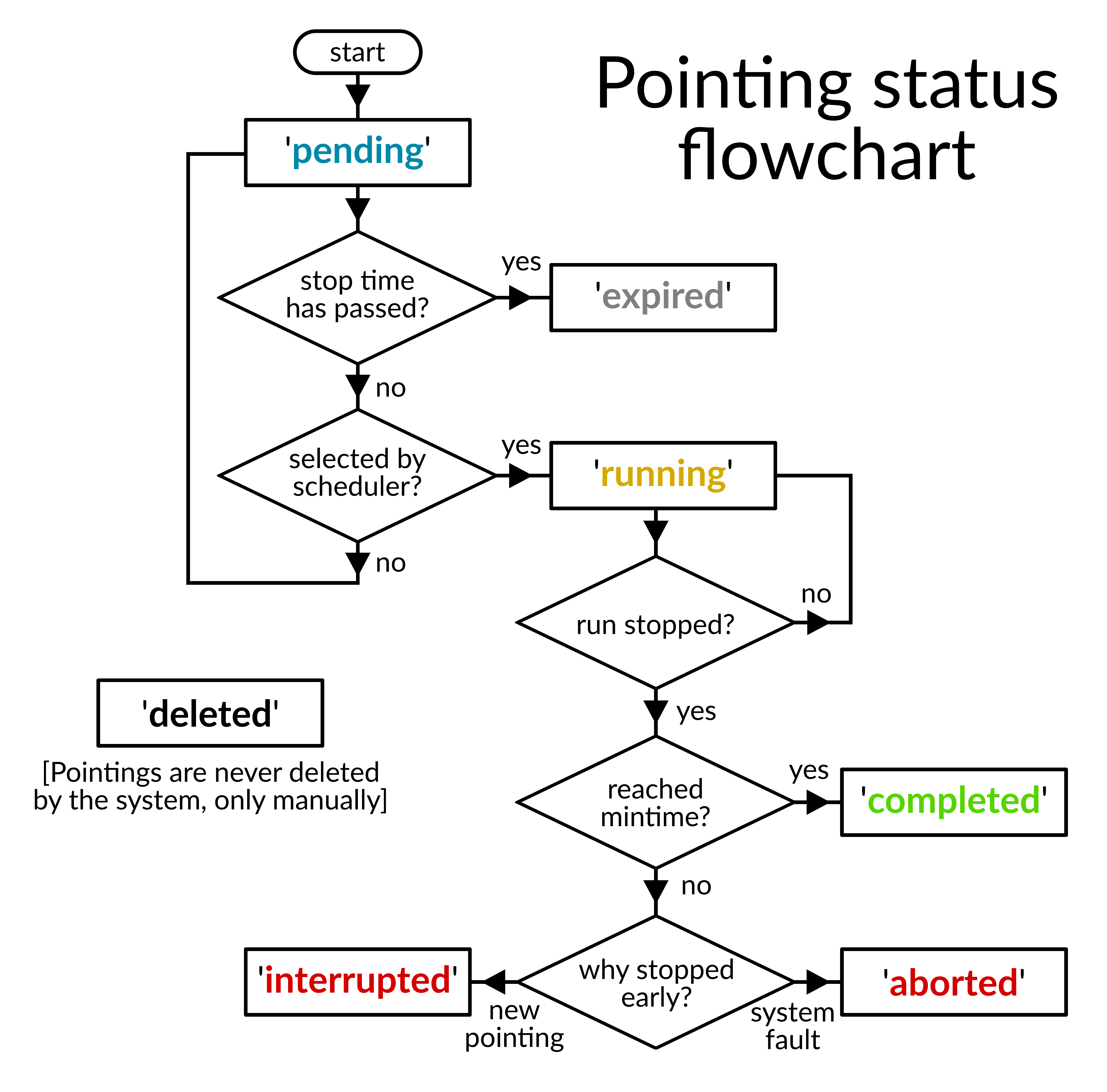}
\end{tabular}
\end{center}
\caption[example] 
{ \label{fig:pointings} 
A flowchart showing how the status of an entry in the pointings table can change.}
\end{figure}

The scheduling system for \textsf{G-TeCS} is based around a MariaDB database, known as the observation database. The primary table in the database is for individual \textit{pointings}. These each represent a single visit of the telescope, with defined RA and Dec coordinates and a valid time range for it to be observed (a start time and a stop time) as well as other parameters.

Each pointing has a status value which is either \textit{pending}, \textit{running}, \textit{completed} or some other terminal status (\textit{aborted}, \textit{interrupted}, \textit{expired} or \textit{deleted}). Ideally a pointing passes through three stages: it is created as pending and sits around in the database, the scheduler eventually selects it and the pilot marks it as running, then if all is well when it is finished it is marked as completed. If it stays in the database and never gets observed it will eventually pass its defined stop time (if it has one) and will be marked from pending to expired by the database caretaker script. If the pointing is in the middle of being observed but is then cancelled before being completed it will be marked either interrupted (if the scheduler decided to observe another pointing of a higher priority) or aborted (in the case of a system problem such as having to close for bad weather). The deleted status is never set automatically, and is reserved for pointings being manually removed from the queue for whatever reason. A representation of the relationship between the pointing status values and how they progress is shown in Fig.~\ref{fig:pointings}.

As well as the target information (RA, Dec, target name) a pointing entry contain limits and constraints about when they can be observed. Each pointing can have a start and stop times set as already mentioned; the scheduler will only select pointings where the current time is within their valid range (and once the stop time has passed they will be marked as expired). Limits can also be set on minimum target altitude, minimum distance from the Moon, maximum Moon brightness (in terms of Bright/Grey/Dark time) and maximum Sun altitude. These constraints are applied by the scheduler to each pointing when deciding which to observe, and unless they all pass the pointing is deemed invalid. When created a pointing is also assigned a rank, usually from 0-9, as well as a True/False flag marking it as a time-critical Target of Opportunity (ToO). These are used when calculating the priority for the pointing, see Sec.~\ref{sec:scheduler}.

The commands to be executed once the telescope has slewed to a pointing are stored in the \textit{exposure\_sets} table. An exposure set defines what commands the pilot will give to the exposure queue daemon. The table has columns for the number of exposures to take, the exposure time and the filter to use. For example, a typical GOTO observation involves taking three 120 second exposures in the L filter followed by one each in the R, G and B filters. This would take up four entries in the pointings table, the first with number of exposures as 3 and filter as L and the rest with number of exposures as 1 and the individual filters. When the pointing is observed by the pilot it will add all linked sets to the exposure queue where each is executed in turn (See Sec.~\ref{sec:focfiltcam}).

Each entry in the pointings table can only be observed once. For observing a target more than once there also exists the \textit{mpointings} table, which contains information to dynamically re-generate pointings for a given target. An mpointing entry is defined with three key values: the requested number of observations, the time each should be valid in the queue and the time to wait between each observation. Each time the database caretaker script is run it looks for any entries in the mpointing table that still have observations to do and it creates another entry in the pointings table for that target. Setting the time values allows a lot of control over when pointings can be valid, for example scheduling follow-up observations a set number of hours or days after an initial pointing is observed.

\begin{figure} [t]
\begin{center}
\begin{tabular}{c}
\includegraphics[width=14cm]{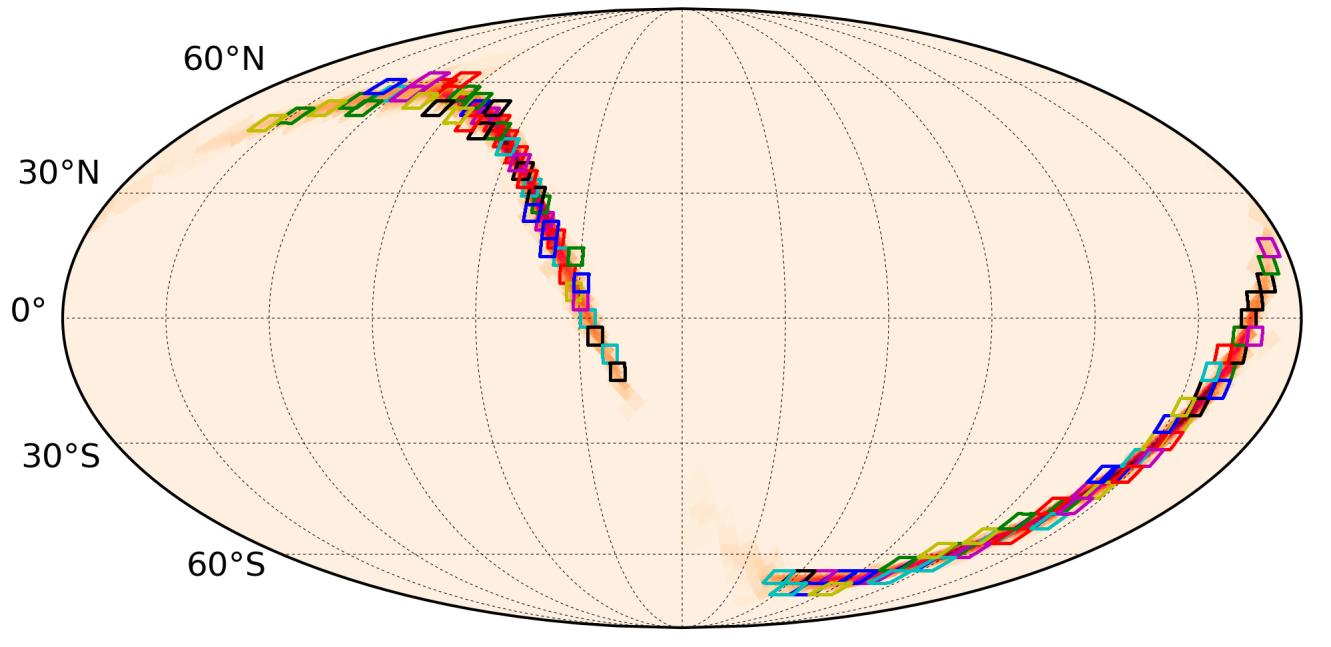}
\end{tabular}
\end{center}
\caption[example] 
{ \label{fig:tiling} 
The probability skymap for gravitational wave event GW151226 \cite{PhysRevLett.116.241103}, covered in tiles matching the field of view of GOTO in the four unit telescope configuration.}
\end{figure}

\subsubsection{Sentinel}
\label{sec:sentinel}

In order for targets to be observed by GOTO in robotic mode they must have entries defined in the pointings (or mpointings) table in the observation database. These can be added manually, using a command line script or a web interface, but for automated observation they have to be inserted whenever an alert is detected. This is the job of the sentinel daemon, as shown in Fig.~\ref{fig:flow}. The sentinel acts as the alert listener for the system, taking any incoming targets and adding them to the database. For individual well-located sources like supernovae a single target is sufficient, but GOTO was designed to cover large areas resulting from gravitational wave skymaps. When an alert with a skymap is detected the sentinel calls functions from a Python module called \textsf{goto-tile}, which maps the skymap onto the fixed GOTO all-sky grid and returns a list of tile pointings and contained skymap probability within each. These are then added to the database by and processed by the scheduler daemon as described in Sec.~\ref{sec:scheduler}. An example of a skymap and the tiles output by \textsf{goto-tile} is shown in Fig.~\ref{fig:tiling}.

\newpage

\subsubsection{Scheduler}
\label{sec:scheduler}

The scheduler daemon runs on the master server that also holds the primary observation database (see Sec.~\ref{sec:obsdb}). There is no specific command loop or hardware restrictions like in other daemons, but there are several advantages to having the scheduling calculations take place in their own daemon on the central server. Firstly, having the calculations done on the same machine as the database is more efficient when it comes to making queries, it is also a more powerful machine than the control computer that the pilot is running on. Furthermore when the second GOTO instrument is built having one scheduler that can speak to either pilot as required opens up further observation strategy opportunities.

The pilot queries the scheduler daemon every 10 seconds, and the job of the daemon is to return the database ID of the pointing that it has calculated is best to observe at the present time. This means GOTO operates under a ``just-in-time'' scheduling model, rather than creating a fixed plan at the beginning of the night of what to observe at each time. This system is very reactive to any incoming alerts as higher-priority pointings will immediately get included naturally in the calculation. The downside is that it less efficient for predefined targets than a night plan, which can be optimised before the night starts. However most of the time GOTO will normally be observing its all-sky survey, which has no strict timing requirements, and any other observations will be alerts entered by the sentinel daemon which could not be planned for.

When called, the scheduler will import the current queue of pointings from the observation database, filter out any that are invalid based on their individual constraints, then calculate a single priority value for each and find the pointing with the highest priority. The scheduling function then returns one of three results: carry on with the current observation, switch to a new observation or park the telescope in the case that there is no valid target. This decision will be returned to the pilot which will either continue what it is doing in the first case, or issue the necessary commands if required. 

The first stage requires the scheduler to find the current queue by filtering the database pointings table. Firstly, pointings are selected based on their defined start and stop times by ensuring the current time is after their start time and before their stop time. Then the pointings are filtered in right ascension and declination by using the current local sidereal time to reject any pointings that would not be visible in the sky. After this initial basic filtering further constraints are applied by the scheduler based on the saved values for each pointing. Pointings have limits defined in Sec.~\ref{sec:obsdb} for physical constraints (altitude, Moon separation, Moon phase, Sun altitude) which are applied using the Constraints system in the \textsf{astroplan} Python module \cite{2018AJ....155..128M}. This ensures that only valid pointings are considered going forward.

The most important part of the scheduling function is calculating the priority value of each valid pointing in the queue. Each pointing entered into the GOTO database will have an assigned rank, which is fixed. The full priority is a floating point number that takes this rank and adds on values for other criteria so it can be easily compared to other pointings. The pointing with the smallest overall priority value is the highest priority to observe. The current criteria used are:

\begin{itemize}

\item The base rank of the pointing ($R$), which is fixed when created. Every pointing is given a rank between 0 and 9, and are deliberately biased towards prioritising gravitational wave events. Ranks 1 to 5 are reserved for these events, with ranks 6 to 9 reserved for other targets, which ensures that any incoming gravitational wave pointings will always be prioritised. Rank 0 is intended to never be used under normal circumstances, but it is a valid value that if set would outrank even gravitational wave events. Survey tiles all automatically have the lowest possible rank of 999, as they are the ever-present background and act as ``queue fillers'' in the system.

\item The repeat number ($N_{r}$), which accounts for the number of times that a particular target
or tile has already been observed, up to 99 times. This ensures that newer pointings are prioritised over later ones. Infinitely-repeating pointings like the survey tiles do not include this, which is why the survey tile ranks are static at 999.

\item The Target of Opportunity True/False flag ($F_{ToO}$). As pointings with the lowest priority value are preferred, pointings marked as ToOs will have $F_{ToO}=0$ and non-ToOs will have $F_{ToO}=1$.

\item The airmass of the pointing at the current time ($X$), calculated by the scheduler each time it is called. Including this will prioritise targets that are at lower airmasses and therefore provide the best quality data. Airmasses between 1 and 3 (altitudes between 90{\textdegree} and 20{\textdegree}) are converted into a value between 0 and 1.
 
\item For gravitational wave or other skymap pointings, the probability contained in the tile associated with the pointing ($P$). This will be calculated by the sentinel using \textsf{goto-tile} and linked to each pointing entered into the database (see Sec.~\ref{sec:sentinel}). The probability is inverted, as priority is ordered by the smallest value, therefore a tile with a 100\% probability of containing the source would have a value of $P=0$ and a tile with a 5\% probability would have $P=0.95$. Pointings that do not have probability values have the value set to 0 by default.

\item For survey tiles, the time since that tile was last observed ($T$). This is also scaled into a value in the range 0 to 1, where 0 corresponds to 7 days or more since the last observation.

\end{itemize}

These values are combined to form a single priority number using the formula:

\begin{equation}
Priority = 10 \times N_r + R + 0.1 \times F_{ToO} + 0.01 \times \left(\frac{a}{a+p+t}X + \frac{p}{a+p+t}P + \frac{t}{a+p+t}T\right),
\label{eq:priority}
\end{equation}

where $a$, $p$ and $t$ are weight factors for the airmass, probability and time-since-observed values. Each pointing in the queue has its priority calculated using Equation~(\ref{eq:priority}), the pointings are then sorted based on priority and the one with the smallest priority value is selected as the highest priority pointing.

\begin{table}[b]
\caption{Some examples of calculated priority values. In this example the weight factors are $a=1$, $p=10$, and $t=10$. The final priority value is constructed using Equation~\ref{eq:priority}. The elements are coloured matching which part of the priority value they form - the rank (\textcolor{red}{red}) and repeat number (\textcolor{orange}{orange}) form the integer part of the priority, the ToO flag (\textcolor{violet}{purple}) forms the first decimal place and the tie-break parameters (airmass, probability and time-since-observed, \textcolor{teal}{teal}) combined form the later decimal places.
}
\label{tab:priority}
\begin{center}
\begin{tabular}{|c|l|c|c|lc|lc|lc|lc|r|}

\hline
~ &
\textbf{Target name} &
\textbf{$R$} &
\textbf{$N_r$} &
\textbf{ToO} &
$F_{ToO}$ &
\textbf{Am.} &
$X$ &
\textbf{Prob.} &
$P$ &
\textbf{Time} &
$T$ &
\textbf{Priority}
\\
\hline

1 &
GW181202 T4 &
\textcolor{red}{1} &
\textcolor{orange}{0} &
True   & \textcolor{violet}{0}   &
1.1    & \textcolor{teal}{0.050} &
4.5\%  & \textcolor{teal}{0.955} &
$-$    & \textcolor{teal}{0}     &
\textcolor{orange}{~~}\textcolor{red}{1}$.$\textcolor{violet}{0}\textcolor{teal}{457} \\

2 &
GW181202 T9 &
\textcolor{red}{1} &
\textcolor{orange}{0} &
True   & \textcolor{violet}{0}   &
1.1    & \textcolor{teal}{0.050} &
0.3\%  & \textcolor{teal}{0.997} &
$-$    & \textcolor{teal}{0}     &
\textcolor{orange}{~~}\textcolor{red}{1}$.$\textcolor{violet}{0}\textcolor{teal}{477} \\

3 &
M31 &
\textcolor{red}{8} &
\textcolor{orange}{0} &
False  & \textcolor{violet}{1}   &
1.0    & \textcolor{teal}{0.000} &
$-$    & \textcolor{teal}{0}     &
$-$    & \textcolor{teal}{0}     &
\textcolor{orange}{~~}\textcolor{red}{8}$.$\textcolor{violet}{1}\textcolor{teal}{000} \\

4 &
GW181202 T3 &
\textcolor{red}{1} &
\textcolor{orange}{1} &
True   & \textcolor{violet}{0}   &
1.1    & \textcolor{teal}{0.050} &
9.1\%  & \textcolor{teal}{0.909} &
$-$    & \textcolor{teal}{0}     &
\textcolor{orange}{~1}\textcolor{red}{1}$.$\textcolor{violet}{0}\textcolor{teal}{435} \\

5 &
AT 2018bdk &
\textcolor{red}{6} &
\textcolor{orange}{2} &
True   & \textcolor{violet}{0}   &
1.0    & \textcolor{teal}{0.000} &
$-$    & \textcolor{teal}{0}     &
$-$    & \textcolor{teal}{0}     &
\textcolor{orange}{~2}\textcolor{red}{6}$.$\textcolor{violet}{0}\textcolor{teal}{000} \\

6 &
AT 2018bfe &
\textcolor{red}{6} &
\textcolor{orange}{2} &
True   & \textcolor{violet}{0}   &
1.2    & \textcolor{teal}{0.100} &
$-$    & \textcolor{teal}{0}     &
$-$    & \textcolor{teal}{0}     &
\textcolor{orange}{~2}\textcolor{red}{6}$.$\textcolor{violet}{0}\textcolor{teal}{005} \\

7 &
M101 &
\textcolor{red}{6} &
\textcolor{orange}{2} &
False  & \textcolor{violet}{1}   &
1.1    & \textcolor{teal}{0.050} &
$-$    & \textcolor{teal}{0}     &
$-$    & \textcolor{teal}{0}     &
\textcolor{orange}{~2}\textcolor{red}{6}$.$\textcolor{violet}{1}\textcolor{teal}{024} \\

8 &
Survey T31 &
\textcolor{red}{999} &
$-$ &
False  & \textcolor{violet}{1}   &
1.0    & \textcolor{teal}{0.000} &
$-$    & \textcolor{teal}{0}     &
4 days & \textcolor{teal}{0.429} &
\textcolor{orange}{}\textcolor{red}{999}$.$\textcolor{violet}{1}\textcolor{teal}{204} \\

9 &
Survey T33 &
\textcolor{red}{999} &
$-$                   &
False  & \textcolor{violet}{1}   &
1.0    & \textcolor{teal}{0.000} &
$-$    & \textcolor{teal}{0}     &
2 days & \textcolor{teal}{0.714} &
\textcolor{orange}{}\textcolor{red}{999}$.$\textcolor{violet}{1}\textcolor{teal}{340} \\

\hline
\end{tabular}
\end{center}
\end{table} 

Some examples of calculated priority values using this method are given in Table~\ref{tab:priority}. The first two pointings are both gravitational wave tiles from the same (fictional) event. They have been inserted at rank 1, neither have been observed yet (meaning $N_r=0$), both are flagged as ToOs and both are at the same airmass. Therefore tie-breaking between the two comes down to the contained probability, and the tile with the higher contained probability (T4) is sorted first. There is another tile from the same event in the queue (T3) with an even higher contained probability, but as it has been marked as already having been observed once ($N_r=1$) is sorted further down below a pointing of M31 manually inserted at rank 8. Next there are three pointings all with the same rank ($R=6$) that have all been observed twice already ($N_r=2$). The first two transient events are both marked as ToOs while the pointing of M101 is not, so it is sorted last of the three. Choosing between the two transient targets comes down to the airmass, with the one currently higher in the sky (so at lower airmass) being sorted first. Finally two tiles from the all-sky survey are shown. These are both automatically fixed at rank 999, and in this case as they are both at the same airmass. Tile 31 is shown as being last observed 4 days ago while Tile 33 was observed only 2 days ago, so Tile 31 comes out with the higher priority.

As demonstrated in this example, priorities can be sorted easily in the first case by considering their repeat numbers, ranks and ToO flags. In the cases these are the same then the tiebreaking value is needed. This is most important when a new gravitational wave event comes in, which can result in up to one hundred new pointings being inserted at once at the same rank. The tiebreaker is a weighted combination of the airmass, contained probability (for skymap tiles) and time since last observation (for survey tiles). The exact weighting is shown in Equation~(\ref{eq:priority}) and depends on the values of $a$, $p$ and $t$. For standard non-skymap, non-survey pointings only airmass is considered, as $P$ is always set to 1 and $T$ is always 0 (as shown in Table~\ref{tab:priority}). Therefore pointings will be selected based on their altitude. For gravitational wave pointings the weighting between airmass and probability determines the observation strategy. Weighting airmass higher will prioritise pointings at a higher altitude and therefore increase typical data quality, but potentially at the loss of GW skymap coverage. On the other hand prioritising only probability would ensure the maximum possible coverage of the skymap, but low-altitude pointings might be selected when it might instead be better to wait until they were at a higher airmass. To attempt to find optimal weighting values many simulations were run with different skymaps and values. These simulations ran through a full night of observing, with the gravitational wave tiles inserted at a random time that day - so simulating both an alert occurring during the night interrupting observations and the alert occurring during daytime so the pointings are there as soon as the system starts observing. Telescope downtime due to weather was also simulated as a Gaussian process with a 10\% chance of occurring and a typical timescale of 1 hour. The initial results of these simulations indicated a weighting of $a=1$ and $p=10$ provided the best optimisation between coverage and average airmass. 

Once every pointing in the queue has had its priority calculated, the pointing with the smallest value is selected as the highest priority to observe. However before returning this pointing's database ID the  scheduler takes into account what the pilot is currently observing. In most cases the pilot will currently be observing a pointing previously given by the scheduler, and on the next check the scheduler will return that the same pointing is still the highest priority - in which case pilot will continue observing. Even if the scheduler finds that a different pointing now has a higher priority it will not tell the pilot to change targets in the middle of observing unless the new pointing has the ToO flag set. Otherwise the pilot will wait until it has finished the current job. Finally if the scheduler finds nothing to do the pilot will park the telescope and wait for the next valid target. 


\section{CONCLUSION}
\label{sec:conclusion}

We have described the custom control and scheduling system developed for the Gravitational-wave Optical Transient Observer. At the time of writing GOTO is still in the commissioning phase and development of the control system is ongoing. The core architecture and hardware control systems are operational, and the system has run for multiple weeks under fully-robotic pilot control. The basic scheduling system has also been effective, but more development of the automated sentinel alert listener is required before the system can be considered fully autonomous.

The addition of the next set of unit telescopes on the first GOTO mount should require very few changes to the control system. However the construction of the second instrument co-located on La Palma opens many more opportunities for advanced scheduling strategy. For example the combined facility could cover the most sky area by always observing different targets with the two instruments, and this will be the default behaviour for the survey observations. However it would be possible to allow the two mounts to both observe the same target at the same time. This could give the option of simultaneous observations in different filters, or allow deeper observations by combining the output of the two instruments.


\acknowledgments 
 
The GOTO Observatory is a collaboration between the University of Warwick and Monash University (as the Monash-Warwick Alliance), Armagh Observatory and Planetarium, the University of Sheffield, the University of Leicester, the National Astronomical Research Institute of Thailand (NARIT), the Instituto de Astrof\'{i}sica de Canarias (IAC), the University of Turku\footnotemark[1] and Rene Breton (University of Manchester)\footnotemark[1].

\footnotetext[1]{membership pending}

\noindent This project makes use of Astropy, a community-developed core Python package for Astronomy\cite{2013A&A...558A..33A,2018arXiv180102634T}.

\newpage
\bibliography{report} 
\bibliographystyle{spiebib} 

\end{document}